\documentclass[a4paper]{article}
\usepackage{amsfonts}
\usepackage{physics}
\usepackage{jheppub}
\usepackage{graphicx}
\usepackage[section]{placeins}
\usepackage{xcolor}
\hypersetup{
  colorlinks=true,
  citecolor=magenta,
  linkcolor=blue,
  urlcolor=violet
 }

\usepackage[utf8]{inputenc}
\usepackage[T1]{fontenc}
\usepackage{lmodern}

\def\be{\begin{equation}}\def\ee{\end{equation}}

\def\md{\mathrm{d}}
\def\mi{\mathrm{i}}

\title{Entanglement Entropy, Relative Entropy and Duality }

\author[a]{Upamanyu Moitra,}
\author[a,b]{Ronak M Soni}
\author[a]{and Sandip P. Trivedi}

\affiliation[a]{Department of Theoretical Physics, Tata Institute of Fundamental Research, Colaba,\\ Mumbai -- 400005, India}
\affiliation[b]{Stanford Institute of Theoretical Physics, Stanford University, Stanford, CA 94305, USA}

\emailAdd{upamanyu@theory.tifr.res.in}
\emailAdd{ronakms@stanford.edu}
\emailAdd{sandip@theory.tifr.res.in}

\preprint{\parbox{3cm}{TIFR/TH/18-23}}

\abstract{A definition for the entanglement entropy in both Abelian and non-Abelian  gauge theories has  been given in the literature, based on an extended Hilbert space construction. The result can be expressed as a sum of two terms, a classical  term and a quantum term. It has been  argued that only the quantum term is extractable through the processes of quantum distillation and  dilution. Here we consider gauge theories in the continuum limit and argue that quite generically, the classical piece  is dominated by  modes with very high  momentum, of order the cut-off,   in the direction normal  to the entangling surface. As a result, we find that the classical term does not contribute to the relative entropy or the mutual information, in the continuum limit,  for states which only carry a finite amount of energy above the ground state. We extend these considerations for $p$-form theories, and also discuss some aspects pertaining to  electric-magnetic duality. }

\begin{document}

\maketitle
\flushbottom

\section{Introduction }\label{intro}

Entanglement entropy is an important way to quantify   quantum correlations in a system.
For gauge theories, the definition of entanglement entropy turns out to be subtle. 
Physically, this is because there are non-local excitations in the system -- for example,  loops of electric or magnetic flux created by Wilson or 't Hooft loop operators,  which can cut across the boundary of the region  of interest. More precisely, it is because the Hilbert space of gauge invariant states, ${\cal H}_{\mathrm{ginv}}$, does not admit a tensor product decomposition between the region of interest and its complement.  

A definition of the entanglement entropy has been  given for a general gauge theory  based on an extended Hilbert space (EHS) construction, see \cite{Buividovich2008,Donnelly2012,Ghosh:2015iwa,Aoki:2015bsa}. For a non-Abelian theory without matter, on a spatial lattice, this  takes the form, 
 \begin{equation}
 \label{resee}
 S_{\mathrm{EE}}=-\sum_i p_i\log p_i +\sum_i p_i \sum_a\log(d^i_a) -\sum_i p_i \Tr_{{\cal H}_i} {\bar \rho}_i \log({\bar \rho}_i).
 \end{equation}
 The  index $i$ in the  sum denotes various sectors which are determined by the value of  the normal component of the  electric field at all the boundary points.
 The probability to be in the sector $i$ is given by $p_i$, and $d^i_a$, with index $a$ denoting the particular boundary point, is the dimension of the representation  specifying the 
 outgoing electric flux at that boundary point in the $i$\textsuperscript{th} sector. The  last term  is a weighted sum with ${\bar \rho}_i$ being the reduced density matrix in the $i$\textsuperscript{th} sector and 
 the trace being taken over the Hilbert space ${\cal H}_i$ in this sector.   It was  argued in \cite{Ghosh:2015iwa,Soni:2016ogt} that this EHS definition agrees with the replica trick method of calculating the entanglement entropy.\footnote{More precisely, the agreement is with a suitably defined replica trick path integral.}  In the discussion below, we will sometimes refer to the first two terms in eq.(\ref{resee}) as the classical terms and the last term as the quantum term.

  For an Abelian theory, the middle term in eq.(\ref{resee}), depending on $d_a^i$, is absent.
The result in the Abelian case agrees with the definition given in \cite{Casini:2013rba}.  More generally, the authors of \cite{Casini:2013rba} argued that the lack of a tensor product decomposition of the Hilbert space was related to the presence of a non-trivial centre. While the Hilbert space of gauge invariant states did not admit a tensor product decomposition, it could be written as a sum of tensor product terms, 
 \begin{equation}
\label{stp}
{\cal H}_{\mathrm{ginv}}=\bigoplus_{i} {\cal H}_i \otimes {\cal H}_i',
\end{equation}
 where each factor ${\cal H}_i \otimes {\cal H}_i'$ is the Hilbert space in a  sector corresponding to a particular value for the centre. 
 Different choices for the algebra of gauge invariant operators  lead to different choices of centres and to a different value of the entanglement entropy in general. 
 In particular, keeping all gauge invariant operators in the region of interest in the algebra gives rise to the electric centre which is specified by the normal component of the electric field along the boundary. It is this definition which agrees with the EHS definition mentioned above for the Abelian case. 
 Other choices of centres can also be made. In particular,  removing all the tangential components of the electric field at the boundary from the algebra gives rise to a magnetic centre (see \cite{Casini:2013rba}).

 In \cite{Soni:2015yga}, (see also \cite{VanAcoleyen:2015ccp}) the EHS definition was compared with a more operational definition based on entanglement distillation and dilution. It was shown that only the last term in eq.(\ref{resee}) corresponds to entanglement that could be distilled into or diluted from a set of Bell pairs using gauge invariant operators that act only in the region of interest or its complement. In contrast, the first two terms do not give rise to any extractable entanglement of this type. 
 In fact, the different sectors, specified by the index $i$, should really be thought of as superselection sectors which cannot be changed by gauge invariant operators acting solely in the region of interest or its complement and, as a result, the probabilities $p_i$ and dimensions $d_a^i$ cannot be changed by these operators. 
 
 In this paper, we are interested in studying the behaviour of the first two terms in eq.(\ref{resee}), which correspond to the non-extractable contributions to the entanglement entropy, in the continuum limit. These terms depend on the  distribution  $p_i$  which determines the probability for being in the various superselection sectors. For the electric centre case,\footnote{In the non-Abelian case the electric centre definition is sometimes  taken to be different from the EHS definition with the middle term in eq.(\ref{resee}) being  absent. We will not be very careful about such distinctions. Our considerations, studying the probability distribution $p_i$, will apply to both  these cases.} by studying the correlation functions of the electric field on the boundary, we will find, both for the Abelian and non-Abelian cases,  
  that this distribution is typically determined by 
 very high momentum modes localised close to the boundary. As a result, we  argue that the  contribution from these terms drops out of the mutual information of disjoint regions or the relative entropy between two states which only have finite energy excitations about the vacuum. 
  
The resulting picture then is quite satisfactory. The extractable part, which contributes to distillation or dilution, is also the only  part  of the entanglement which 
contributes to the mutual information or the relative entropy. 

It should also be noted here that \cite{Casini:2013rba} had used an inclusion argument, together with the monotonicity of relative entropy (see \cite{ohya2004quantum,Witten:2018lha} for reviews), to argue that the dependence on the choice of centres should drop out in the continuum limit. This ties in nicely with our result above pertaining to the electric centre case which is obtained by keeping all the gauge invariant operators in the region of interest in the algebra. Together, the two results imply that, in the continuum limit, regardless of the choice of centre, the relative entropy and mutual information only get contributions from the entanglement which can be extracted using all gauge invariant operators acting on the region or its complement.

 Two additional aspects are also discussed in this paper. 
 
 We extend the discussion of the probability distribution $p_i$ for the electric centre case to general $p$-form theories. For the usual gauge theory ($p=1$) in $d+1$ dimensions, the probability distribution is determined by the Laplacian for a scalar that lives on the $(d-1)$-dimensional spatial boundary of the region \cite{Eling2013,Huang2015,Donnelly:2014fua,Donnelly2016,Zuo2016,Soni:2016ogt}. By considering a planar boundary, we argue that the result for a $p$-form is determined by the Laplacian operator  for a $(p-1)$-form living on the $(d-1)$-dimensional spatial boundary.  This ties in with the earlier analysis in \cite{Donnelly:2016mlc,Dowker:2017flz}. 
 We also discuss some aspects of electric-magnetic duality which exchanges a $p$-form theory with a $(d-p-1)$-form theory in $d+1$ spacetime dimensions. 
 Working on a spatial lattice, we argue that the magnetic centre of the $p$-form theory maps to the electric centre in the dual theory with the addition of a few more operators in the algebra. In the continuum limit, these extra operators will not make a difference in computing the relative entropy or the mutual information. 
 
 This paper is structured as follows. The classical term and, in particular, the probability $p_i$ on which it depends are discussed in section \ref{sec:classical}.
 Following that, we discuss some aspects of duality in section \ref{sec-duality}. We end with some conclusions in section \ref{sec:conclu}.

 Before proceeding, let us also point out some of the other relevant and growing literature on the subject: \cite{Gromov:2014kia,Radicevic:2014kqa,Donnelly:2014gva,Casini:2014aia,Hung:2015fla,Radicevic:2015sza,Ma:2015xes,Itou:2015cyu,Casini:2015dsg,Radicevic:2016tlt,Nozaki:2016mcy,Aoki:2016lma,Balasubramanian:2016xho,Agarwal:2016cir,Aoki:2017ntc,Hategan:2017jts,Hung:2017jnh,Pretko:2018nsz,Blommaert:2018rsf,Hategan:2018uas,Anber:2018ohz,Blommaert:2018oue,Lin:2018bud,Huerta:2018xvl}.

\section{Classical Term in the Continuum limit}\label{sec:classical}
\subsection{$U(1)$ Abelian Theory}\label{subsec:abel}
Let us begin by considering a $3+1$-dimensional free $U(1)$ theory. The entanglement entropy in this theory for a spherical region of radius $R$ was studied in \cite{Eling2013,Huang2015,Donnelly:2014fua,Donnelly2016,Zuo2016,Soni:2016ogt}.
Since the theory is free, the classical term is determined by the two-point function of the normal component of the electric field,
\begin{equation}
\label{defy}
G_{rr}=\langle E_r E_r\rangle ,
\end{equation}
with the  two  electric fields $E_r$ being  inserted at two different points on the boundary which is an  $S^2$ of radius $R$. The classical term is then given by 
 \begin{equation}
 \label{classy}
 -\sum_i p_i \log(p_i) = \alpha_1  {A \over \epsilon^2} -{1\over 2} \log(\det G_{rr}^{-1}),
 \end{equation}
 with $\epsilon$ being  a short distance cut-off. Thus, up to  a non-universal area law divergence, $G_{rr}$ determines the classical term. Let us note in passing that the first term on the RHS of eq.(\ref{classy}) depends on the measure for the sum over all electric field configurations. Starting from the lattice and passing to the continuum gives a well defined measure as discussed in \cite{Soni:2016ogt}.

 Since the two electric fields in $G_{rr}$, eq.(\ref{classy}),    are inserted at points on the boundary $S^2$, as was mentioned above,   it is a function of the angular separation of the two points. 
 In fact, it turns out to be divergent and some care needs to be exercised in regulating this divergence and defining it. 
 After introducing a short distance cut-off $\Delta$  for the radial momentum, (which is the component of the momentum along the  direction normal to the boundary $S^2$) the resulting answer was shown to be \cite{Soni:2016ogt}, 
 \begin{equation}
 \label{unsp}
 G_{rr}^{lm}={1\over \pi R^4} \left( \log {R^2\over \Delta^2} \right) l (l+1). 
 \end{equation}
 Here we have carried out a Fourier transform to go from the angular separation variables  to the angular momentum variables labelled by integers $(l,m)$, as per standard conventions. 
 Note that the result diverges as $\Delta \rightarrow 0$, due to the contribution of modes with very high values of the radial momentum, which dominate the result in this limit. 
 The factor of $l (l+1)$ means that the Green function is  proportional to the two-dimensional Laplacian for a free scalar on the boundary.

 From $G_{rr}$, we can calculate the classical piece, eq.(\ref{classy}) which  turns out to be, 
 \begin{equation}
 \label{classfi}
 -\sum_ip_i\log(p_i)= \alpha_2 {A \over \epsilon^2} -{1\over 6} \log ({A\over \epsilon^2}) + \cdots,
 \end{equation}
 where the ellipsis refers to non-universal finite pieces. 
 The logarithmic term arises from the two-dimensional scalar Laplacian mentioned above, while the pre-factor $  \log(R^2/\Delta^2)/\pi R^4$ contributes to the non-universal area law divergent piece above. 
 
 It is important to emphasise that we have introduced two cut-offs  above, $\Delta$ and $\epsilon$, and worked in the limit where $\Delta$ goes to zero first.\footnote{We thank W. Donnelly for a discussion on this point.} 
 This was true in eq.(\ref{unsp}), for example, since the angular momentum $l$ along the $S^2$ was being kept fixed while $\Delta\rightarrow 0$. 
 It is in this limit that $G_{rr}$ depends on the two-dimensional Laplacian for a free scalar with a logarithmic dependence on the radial cut-off $\Delta$. 
 The behaviour of modes whose momentum along $S^2$ is comparable to the radial cutoff is more complicated and less universal. 
 In the discussion below, when we analyse the contribution of the classical term to the mutual information and relative entropy, we will comment on such modes as well and show that their contribution too drops out from these quantities.

 Having understood the case of a spherical region, we can now turn to the more general situation.  Let us begin by first considering the half space $z<0$, with the remaining spatial coordinates, $(x,y)$ taking values in $(-\infty, \infty)$. The boundary of this region is 
 the two-plane $z=0$. The normal component of the electric field is along the $z$ direction. 
 It is easy to see by standard quantisation of the electromagnetic field that 
 \begin{equation}
 \label{elf}
 \langle E_z( \mathbf{x}_1) E_z( \mathbf{x}_2) \rangle =  \int {\md^3k \over (2\pi)^3} {\omega_k\over 2} \pqty{ 1-{k_z^2\over k^2}} e^{i \mathbf{k} \cdot (\mathbf{x}_1-\mathbf{x}_2)}.
 \end{equation}
 (The two-point function is computed at equal time.)  Here, $\omega_k = k=\sqrt{k_z^2+k_x^2+k_y^2}$. 
 Setting the two points to be at the boundary, $z=0$, we get 
 \begin{equation}
 \label{twopaa}
 \langle E_z(\mathbf{k}_\parallel) E_z(\mathbf{k}_\parallel') \rangle 
 =   (2\pi)^2 \delta^{(2)}( \mathbf{k}_\parallel + \mathbf{k}_\parallel')\, k_\parallel^2 \int {\md k_z\over 4\pi} {1\over k}.
 \end{equation}
 We see that the integral on the RHS is logarithmically  divergent since $k \sim k_z$ for large $k_z$. Thus, as noted above, the result needs to be regulated by introducing a cut-off for momentum along the $z$ direction, normal to the boundary. One efficient way to do this is to take the two points located at slightly different values in the $z$ direction, with $\Delta z=\Delta$, instead of both of them being exactly at the boundary,  $z=0$.  
 Repeating the calculation above now gives, 
 \begin{equation}
 \label{twopab}
 \langle E_z(\mathbf{k}_\parallel) E_z(\mathbf{k}_\parallel')\rangle 
 =  (2\pi)^2\delta^{(2)} (\mathbf{k}_\parallel + \mathbf{k}_\parallel') \, k_\parallel^2
\int {\md k_z\over  4\pi}{e^{\mi k_z \Delta} \over k}.
\end{equation}
We see that the divergence in the integral at large $k_z$ is  regulated resulting in 
\begin{equation}
\label{twopac}
\langle E_z(\mathbf{k}_\parallel) E_z(\mathbf{k}_\parallel')\rangle 
 \sim   (2\pi)^2\delta^{(2)}( \mathbf{k}_\parallel + \mathbf{k}_\parallel') \, k_\parallel^2 \log({k_\parallel \Delta}).
\end{equation}

Comparing with eq.(\ref{unsp}), we see that the result above is analogous to what was obtained for the spherical region case.  In particular, the result diverges logarithmically as $\Delta \rightarrow 0$  and is proportional to  $k_\parallel^2$  which is the eigenvalue  of the two-dimensional scalar Laplacian on the boundary. One difference is that in the spherical case, the logarithmic divergence due to the modes with high value of the normal component of momentum is cut-off by the radius $R$, whereas in the infinite plane case, where this cut-off is not available, it is cut-off 
by $k_\parallel$.

From this example of the half plane and the spherical region, it is now clear that we expect for the two-point function in any compact region a  result  analogous to the sphere case, namely  going like $\log({R/\Delta} )$,  with $\Delta $ being the cut-off for the normal component of momentum, and $R$ being an IR scale provided by the size of the region of interest, and also being proportional to the two-dimensional Laplacian along the boundary. Additional arguments in support of this were also given in \cite{Eling2013,Huang2015,Donnelly:2014fua,Donnelly2016,Zuo2016,Soni:2016ogt}.

 We are now ready to consider the mutual information and relative entropy in this theory. 
 Consider the mutual information in the vacuum state first. Suppose there are two compact disjoint spatial regions $A, B$. The mutual information is given by 
 \begin{equation}
 \label{mina}
 I(A,B)=S_A+S_B-S_{A B},
 \end{equation}
 where $S_{AB}$ is the entanglement between the region $A\cup B$ and the rest. 
 We are interested, in particular, in the classical term's contribution to $I(A,B)$. Since the $U(1)$ theory under consideration is free, this is determined by the two-point function, as discussed above, with $p(E_n)$ the probability to be in the sector where the normal component of the electric field takes value $E_n$ being given by, 
 \begin{equation}
 \label{valnp}
 p(E_{n}(\mathbf{x}) )=N \exp(-{1\over 2} \int \md^2x \, \md^2 y\, E_{n}({\mathbf x})  G_{nn}^{-1}(\mathbf{x}-\mathbf{y})E_{n}(\mathbf{y})).
 \end{equation}
 Here $N$ is a normalisation, $E_n$ denotes the normal component of the electric field, ${\bf x, y}$ are two points on the boundary and $G_{nn}$ is the Green function for the normal component $E_n$. 
 
 When we are dealing with the two disjoint regions $A$ and $B$, we have two  boundaries which are also disjoint. Thus the two-point function which appears on the RHS is evaluated when the two points ${\bf x,y}$ are both on the boundary of $A$ or on the boundary of $B$, or when one point is on the boundary of $A$ and the other on the boundary of $B$. We saw on general grounds in the previous subsection that when the two points are on the same boundary the two-point function diverges and, after a cut-off is introduced, is  proportional to 
 $\log({R/\Delta})$, where $R$ is an IR scale set by the size of the region  $A$ or $B$. There is no such divergence when the two points are located on the two different boundaries of $A$ and $B$ respectively, since in  this case the two points  cannot come closer to each other.  Hence, the two-point function which appears in the exponent in eq.(\ref{valnp})  will be dominated by the contribution when the two points are on the same boundary; with the contribution from  when they are on separate boundaries being suppressed parametrically by a factor of $[\log(R/\Delta)]^{-1}$. As a result, the probability, $p(E_n^A, E_n^B)$, for the $E_n$ to take the value $E_n^{A}, E_n^{B}$,  on the two boundaries $\partial A, \partial B$ will  simply be  the product, 
 \begin{equation}
 \label{prodp}
 p(E_n^A, E_n^B)=p(E_n^A) p(E_n^B),
 \end{equation}
 where $p(E_n^A)$,   for example,  is the probability that $E_n$ takes value $E_n^A$ on  $\partial A$ regardless of any value  $E_n$ takes  on $\partial B$. Eq.(\ref{prodp}) is true up to subleading terms which vanish in the continuum limit when $\Delta/R\rightarrow 0$.  It then follows that the classical term will cancel out and not contribute to the mutual information in the continuum limit.

 It is worth considering a simple example to explain this point more clearly. 
 Take a  two variable probability distribution with the 
  non-zero moments being
\begin{align}
  \langle x^{2}  \rangle = \langle y^{2} \rangle &= \frac{1}{\epsilon} \gg 1 \nonumber\\
  \langle x y \rangle &= 1.
  \label{eqn:eg-moments}
\end{align}
The probability distribution that gives rise to these moments is
\begin{equation}
\label{proba}
  p(x,y) = N    \exp [-{ \left( {\epsilon \over 2 (1-\epsilon^2)}(x^2+y^2) - {\epsilon^2\over (1-\epsilon^2)} xy \right)} ].
  \end{equation}  
  As one can see, the off-diagonal term is suppressed by a factor of $\epsilon$ compared to the diagonal one. From eq.(\ref{proba}) it follows that 
  \begin{equation}
  \label{probb}
  p(x,y)=p(x)p(y)[1+ \epsilon^2 xy ]+ \cdots,
  \end{equation}
 where the ellipsis denotes terms which are higher order in $\epsilon$. 
 Hence, we see that the probability distribution factorises up to small corrections.
 
 Comparing with eq.(\ref{twopac}), we see that for the gauge theory case of interest, the role of $\epsilon$ is being played by $({\log\Delta})^{-1}$.  Thus the corrections to eq.(\ref{prodp}) are of fractional order $({\log\Delta})^{-2}$, and the mutual information will receive corrections of this order which will vanish when $\Delta\rightarrow 0$.

 Similarly, we can consider the relative entropy between two states. Here we will be interested in states which only carry a finite energy above the ground state.  Let the corresponding density matrices for some spatial region be $\rho^1$ and $\rho^2$ in these two cases; the relative entropy is then given by 
 \begin{equation}
 \label{reledef}
 S(\rho^1|\rho^2)=\Tr(\rho^1 \log \rho^1) - \Tr(\rho^1\log\rho^2).
 \end{equation}
 It is easy to see that this becomes
 \begin{equation}
 \label{reldef2}
 S(\rho^1|\rho^2)= \sum_ip^1_i \log({p^1_i\over p^2_i})+\sum_ip^1_i \Tr_{{\cal H}_i}\log({{\bar \rho_i}^1\over {\bar \rho_i}^2}),
 \end{equation}
 where ${\bar \rho}_i^{1,2}$ are the normalised density matrices in the $i^{th}$ sector in state $1,2$ respectively. 
 We can see, by an argument analogous to the one above for mutual information that the first term above, which is due to the classical contribution to the entanglement, again vanishes in the continuum limit. The argument is as follows. The probability $p^{1,2}_i$ is determined by the correlations functions of $E_n$ in state $1,2$. The two-point function in the two states to the leading order will be the same, and in turn, equal to that in the vacuum, eq.(\ref{twopac}), since it is dominated by very high momentum modes whose behaviour will be the same as in the vacuum for states which only carry a finite energy above the vacuum.  For a region of size $R$, this two-point function goes like $\log({R/\Delta})$ and therefore diverges when $\Delta \rightarrow 0$. Connected higher point  correlations can arise in states which are not the vacuum, but these will be finite and thus, subdominant compared to the two-point function. 
 Therefore, $p^1_i, p^2_i$ will be the same up to corrections. From the discussion above after eq.(\ref{probb}) it follows that 
 
 \begin{equation}
 \label{valrat}
 \log({p^1_i\over p^2_i})\sim 
\left( {1\over \log({R \over \Delta})} \right)^2.
 \end{equation}
 Since $p^1_i$ is normalised so that $\sum_i p^1_i=1$, we see that the first term will be of order ${[\log({R/\Delta})]^{-2}}$ and will thus vanish. 
 
 A few comments are worth making before we proceed. We had mentioned above that,  in general, the two-point function would depend on the boundary Laplacian. For a spherical region, it  was shown in \cite{Soni:2016ogt} and \cite{Donnelly:2016mlc} that the zero mode of the Laplacian needs to be  excluded when computing the determinant,
 due to the Gauss law constraint. More generally, also the zero modes need to be handled with care, but we will not be precise about this here. 
 See \cite{Donnelly:2016mlc} for a careful discussion in this regard. 
 
 The divergent behaviour of the two-point function  going like $\log(R/\Delta)$ is true for modes which carry  momentum along the boundary that is much smaller that $\Delta^{-1}$, (i.e.,   $k_\parallel \ll \Delta^{-1}$).  We can also consider modes which have $k_\parallel\sim \Delta^{-1}$. In this case also the two-point function is dominated by the contribution
  when the two points lie on the same boundary. Consider,  for example, two regions corresponding to $z<0$ and $z>L$. When the two points are on the two boundaries, $z=0, z=L$ we get from eq.(\ref{elf}) that 
  \begin{equation}
  \label{addtwo}
   \langle E_z(\mathbf{k}_\parallel) E_z(\mathbf{k}_\parallel')\rangle 
 =  (2\pi)^2\delta^{(2)} (\mathbf{k}_\parallel + \mathbf{k}_\parallel')\, k_\parallel^2
\int {\md k_z\over  4\pi}{e^{\mi k_z L} \over \sqrt{k_\parallel^2 +k_z^2}}.
  \end{equation}
  Carrying out the integral gives 
  \begin{equation}
  \label{intd}
   \langle E_z(\mathbf{k}_\parallel) E_z(\mathbf{k}_\parallel')\rangle 
 =  2\pi \delta^{(2)} (\mathbf{k}_\parallel + \mathbf{k}_\parallel')\, k_\parallel^2 K_0 ( k_\parallel L),
  \end{equation}
  where $K_0$ is the modified Bessel function of the second kind. 
  For $k_\parallel \sim \Delta^{-1} \gg L^{-1}$, we get 
  \begin{equation}
  \label{intdd}
  \langle E_z(\mathbf{k}_\parallel) E_z(\mathbf{k}_\parallel')\rangle \sim   \delta^{(2)} (\mathbf{k}_\parallel + \mathbf{k}_\parallel')\, \pqty{1 \over \Delta^3 L}^{1/2} e^{-L/\Delta},
  \end{equation}
  so that the two-point function, where the two points lie on the two boundaries for modes with $k_\parallel \sim \Delta^{-1}$   is exponentially suppressed, compared to the case when the two points lie on the same boundary. Thus the contribution of these modes will also drop out in the classical part of the  mutual information. Similarly, if we are considering two states 
  whose behaviour at the cut-off scale $\Delta$ is the same as that of the vacuum, then the contribution of modes with $k_\parallel \sim { \Delta^{-1}}$ will also drop out in the relative entropy of these two states.

  Let us also briefly consider the case of the magnetic centre. In this case the different superselection sectors are specified by the 
 normal component of the magnetic field, ${\mathbf B}$, and the probability of being in a superselection sector is specified by the two-point function 
 of the normal component of ${\mathbf B}$.  
  For the planar boundary considered above at $z=0$, it is easy to see that the two-point function is given by,
  \begin{equation}\label{mag-two-pt}
\langle B_{z} (\mathbf{k}_\parallel) B_{z} (\mathbf{k}'_\parallel) \rangle = (2\pi)^2 \delta^{(2)} ( \mathbf{k}_\parallel+\mathbf{k}'_\parallel) \, {k}_\parallel^2 \int \frac{\md k_z}{4\pi} {e^{\mi k_z \Delta}\over k}.
\end{equation}
This  is the same as  the two-point function for the electric field,  eq.(\ref{twopab}). 

 It is also straightforward to generalise this discussion for a gauge field to other  dimensions. In $d+1$ dimensions, eq.(\ref{elf}) is  replaced by 
 \begin{equation}
 \label{repa}
 \langle E_z(\mathbf{x}_1) E_z(\mathbf{x}_2) \rangle = \int {\md^dk \over (2\pi)^d} {\omega_k \over 2} \pqty{ 1-{k_z^2\over k^2} } e^{\mi \mathbf{k}\cdot (\mathbf{x}_1-\mathbf{x}_2)},
 \end{equation}
 where we are still considering the region $z<0$ with a boundary which now  has extent in $d-1$ spatial dimensions. 
 
 It then follows that
  
\begin{align}
 \label{twopaba}
 \langle E_z(\mathbf{k}_\parallel) E_z(\mathbf{k}_\parallel')\rangle &
 =   (2\pi)^{d-1} \delta^{(d-1)} ( \mathbf k_\parallel + \mathbf k_\parallel' ) \, k_\parallel^2
\int {\md k_z\over 4\pi} {e^{\mi k_z \Delta} \over k} \nonumber \\
&\sim    (2\pi)^{d-1}\delta^{(d-1)} ( \mathbf k_\parallel + \mathbf k_\parallel') \, k_\parallel^2 \log(k_\parallel \Delta),
\end{align}
showing that the logarithmic divergence, and boundary Laplacian are   universal features in all dimensions $d\ge 2$.\footnote{In $d=2$, despite the fact that the compact theory is confined on large length-scales \cite{Polyakov:1975rs}, it is not on small length scales, and thus we expect this result to continue to hold there as well.} The logarithmic divergence then implies that our arguments go through for the $U(1)$ theory in any dimensions $d\ge 2$. It also follows that   the  mutual information and relative entropy  are independent of the classical term.\footnote{Using arguments analogous to eq.(\ref{addtwo})-eq.(\ref{intdd}), it also follows that the contribution from modes with $k_\parallel \sim {\Delta^{-1} }$ to the classical term in the mutual information and relative entropy vanish.} 
 
 These considerations can be extended to $p$-forms and their duals in general dimensions easily, as we will discuss later.

 Next, let us consider adding charged matter to the system. On the lattice, the charged matter would add degrees of freedom that live on the sites of the spatial lattice in the standard manner. The discussion above can be readily extended to such a case as well. The Gauss law constraint  still results in a non-trivial centre, and  the full Hilbert space  admits a decomposition of the form given in eq.(\ref{stp}), where the label $i$ denotes sectors where the centre takes a fixed value. Also, the extended Hilbert space in this case is obtained by taking the tensor product of the Hilbert spaces of the gauge degrees of freedom living on the links and the matter degrees of freedom living on the sites. Passing to the continuum in the electric centre case, or equivalently, the extended Hilbert space case, the different sectors are specified by the different values the normal component of the electric field takes on the boundary of the region of interest, and  $p_i[E_n]$ is a functional of this boundary value. 
 The theory is no longer free and thus, there are higher point correlations of $E_n$, besides the two-point function. 
 
 As long as the interactions become weak in the UV, at the scale of the lattice, one would expect that the theory is perturbative at this scale and the divergence seen in the free field case would dominate over the perturbative corrections due to the interactions. Thus the leading correlation would be the two-point function which goes like eq.(\ref{twopac}) and diverges in the continuum limit. This should  be the case for two spatial dimensions,  $d=2$,  since in that case the gauge coupling is super-renormalisable. In $d=3$, the gauge coupling grows logarithmically and becomes strong in the UV;  it also becomes strong in the UV in   $d>3$ where the gauge coupling is non-renormalisable. In these cases, therefore, it is not clear what happens in the continuum limit or even whether such a limit exists. One interesting possibility is that the theory flows to a non-trivial fixed point in the UV. In this case, the behaviour of the two-point function
 of the electric field would be  determined by its anomalous dimension at the UV  fixed point with the  short distance contribution taking the form
 \begin{equation}
 \label{shdi}
  \langle E_z(\mathbf{k}_\parallel) E_z(\mathbf{k}_\parallel')\rangle 
 \sim   \delta^{(d-1)} ( \mathbf{k}_\parallel + \mathbf{k}_\parallel') \, k_\parallel^2
\int {\md k_z\over k^{(d-2(\delta-1)) }}e^{\mi k_z \Delta}, 
\end{equation} 
 where $\delta$ is the anomalous dimension. If $\delta>{(d+1)/2}$, so that the anomalous dimensions  exceeds the engineering dimension, then the correlations will be even more strongly dominated by the short distance modes and one expects the arguments for the independence of the mutual information and relative entropy  from the classical term to apply. For $\delta <{(d+1)/2}$, the situation is less clear. If the correlations of $E_n$ are  finite and non-divergent,  then it could be    that  the classical piece continues to contribute in the continuum limit to the mutual information and the relative entropy.

 Let us end this subsection with two comments. First, suppose that  the theory is strongly coupled in the UV and flows to a weakly coupled one,  which is close to the free non-interacting theory,  in the IR at an energy scale of order $E \sim \Lambda$. In such cases, one would  still  expect that the contribution to the classical term from modes with energies $E \le \Lambda$ is approximately  governed by the two-point function discussed above and thus, the contribution of these modes should  drop out in the mutual information or the relative entropy. It will be  worth trying to make this intuitive argument more precise \cite{Ghosh:2017gtw}.  Second, as has been mentioned above, we have not been very careful about zero modes of the boundary Laplacian. It is worth pointing out that these can sometimes lead to a non-vanishing contribution from the classical term for the mutual information or relative entropy\footnote{We thank the referee for pointing this out.}.
 Consider, for example, pure Maxwell theory in, say, $3+1$ dimensions when all the spatial directions  $x,y,z$ are compact circles -- and consider two entangling regions which extend along the $x,y$ circles and extend from $0<z<\Delta_1$ and $\Delta_2 <z<\Delta_3$, respectively. 
 This situation can be analysed by dimensionally reducing  along $x,y$  to the $1+1$-dimensional theory in $t,z$. The electric field in this $1+1$-dimensional theory, $E_z$, is spatially constant due to Gauss' law, but need not vanish, since $z$ is  compact. In general, the state of the system will therefore be a linear superposition of different eigenstates of $E_z$, and as a result, a classical contribution will arise in the mutual information which is non-vanishing. Similarly, if we consider one such region and two states which are different linear superpositions of $E_z$ eigenstates, there will be a non-vanishing contribution to the relative entropy. 
 When the $z$ direction is non-compact though, $E_z$ must vanish to keep the energy finite and such a contribution to the mutual information or relative entropy will not arise.

 \subsection{Non-Abelian Theory}\label{subsec:nonabel}
 Let us now turn to the non-Abelian case. In this case, in the continuum, the different sectors in the electric centre or the EHS definition  are  specified by the value of $\Tr(E_n^2)$ and other  Casimirs (i.e., appropriate local gauge invariant operators) on the boundary, where $E_n$ is the normal component of the electric  field. The probability to be in a particular sector is a functional of $\Tr(E_n^2)$ and the other Casimirs and we  denote it for ease of notation as $p[E_n^2]$. 
 
 Let us start by considering the free $SU(N) $ Yang Mills theory, say, in $3+1$ dimensions, and consider the region of space $z<0$. The boundary lies at $z=0$. 
 It is easy to show that in this case the two-point function is given by
 \begin{eqnarray}
  \langle E_z(\mathbf{k}_\parallel)^2 E_z(\mathbf{k}_\parallel')^2 \rangle & \sim &  (N^2-1)  \delta^{(2)} ( \mathbf k_\parallel+\mathbf k_\parallel') 
  \int \md^2{\tilde k_\parallel}\,{\tilde k_\parallel}^2( \mathbf{k}_\parallel-{\tilde{\mathbf{k}}_\parallel})^2  \nonumber \\
   &&\int {\md k_z\over \sqrt{k_z^2+ {\tilde k_\parallel}^2}} e^{\mi k_z \Delta} 
   \int {\md k_z'\over \sqrt{k_z'^2 + (\mathbf{k}_\parallel-{\tilde{\mathbf{k}}_\parallel})^2}}e^{\mi k_z'\Delta},
   \end{eqnarray}
 where we have separated the two points in the $z$ direction by a value $\Delta z= \Delta$. 
 We see that now the integral over ${\tilde k}_\parallel$, the momentum along the boundary, is also divergent, since $E_z^2$ is a more singular operator that $E_z$. 
 Introducing a cut-off along the boundary directions and not distinguishing it from $\Delta$, the cut-off along the normal direction,  we get 
 \begin{equation}
 \label{tptym}
 \langle E_z(\mathbf{k}_\parallel)^2 E_z(\mathbf{k}_\parallel')^2 \rangle  \sim   (N^2-1) \delta^{(2)}(\mathbf{k}_\parallel+\mathbf{k}_\parallel'){1\over \Delta^6},
 \end{equation}
 up to logarithmic corrections, which we have not kept carefully. 
 This is much more singular than the two-point function of $E_z$ in the Abelian case. 
 
 It is also easy to calculate the higher point correlators for $E_z^2$ in the free case.  The contributions to the $n$-point function can be classified in terms of the number of disconnected components.  The dominant contribution comes from the  maximally disconnected component going like the product $\langle E_z^2(1) E_z^2(2)\rangle \langle E_z^2(3) E_z^2(4)\rangle \cdots \langle E_z^2(n-1) E_z^2(n)\rangle$ (for $n$ even), which diverges as $\Delta^{-2(n+1)}$. In contrast, the fully connected contribution, for example, goes like
 $\Delta^{-6}$, as before, and is therefore subdominant. As a result, the two-point function dominates the correlation functions and hence the probability distribution  and therefore the classical term is determined by the two-point function eq.(\ref{tptym}), which is dominated by the UV modes at the cut-off. 
As for the higher-order Casimirs, the two-point function is even more singular and therefore dominates even more strongly.

 On turning on the gauge coupling, these correlations will change. However, Yang-Mills theory is known to be asymptotically free in $3+1$ dimensions and  super-renormalisable in $2+1$ dimensions. Therefore, in these cases, we expect the behaviour at the scale of the lattice to continue to be that of the free theory to good approximation, and   the two-point and higher point correlations to diverge, as described  above. 
 These divergences of course occur because points on the boundary can come close together. When we consider two disconnected boundaries, this means that the correlations will be dominated by their value when the points lie on the same boundary. As a result, in the continuum limit, the joint probability should satisfy the condition 
 \begin{equation}
 \label{jprob}
 p\pqty{(E_n^2)^A , (E_n^2)^B}=p\pqty{(E_n^2)^A} p \pqty{(E_n^2)^B},
 \end{equation}
 analogous to eq.(\ref{prodp}) in the Abelian case,  
 and the contribution  of the classical piece to the mutual information should therefore  vanish. Note that this conclusion is equally true in the electric centre definition, where the classical piece is given by the first term on the RHS of eq.(\ref{resee}), and in the EHS definition, where it is given by the sum of the first two terms on the RHS of eq.(\ref{resee}).  Both these terms will vanish if eq.(\ref{jprob}) is true. Similarly, one can argue that the relative entropy for states which only differ from the vacuum  with  finite energy excitations, will also not receive a contribution from the classical piece.

 In higher dimension $d>3$, the theory is non-renormalisable and the situation is more interesting. A continuum limit might exist with the theory flowing to a non-trivial fixed  point in the UV, and as in the discussion above for the Abelian case, the anomalous dimension for $\Tr(E^2)$ would then determine the short distance nature of the two-point and higher point correlators. 
 Similarly, adding matter can change the behaviour of the theory. An important example of this type is  if the resulting theory becomes conformally invariant. Once again, the 
 anomalous dimensions for $\Tr(E^2)$ plays an important role in  determining the nature of the short distance correlators. If the correlators are smooth at short distances, and   do not diverge as $\Delta\rightarrow 0$, then the classical piece could potentially  contribute to both the mutual information and the relative entropy. 

\subsection{$p$-Form Abelian Gauge Theory}\label{subsec:pform} 
Our discussion of the Abelian $U(1)$ gauge theory can be easily generalised to the case of a general $p$-form Abelian theory in $d+1$ dimensions. 

The action is given by 
\begin{equation}
\label{p-form-act} 
S= -\frac{1}{2(p+1)!} \int \md^{d+1} x \, H_{\mu_1\mu_2 \cdots \mu_{p+1}} H^{\mu_1\mu_2 \cdots \mu_{p+1}},
\end{equation}
where 
\begin{equation}
\label{defH}
H_{\mu_1\mu_2 \cdots \mu_{p+1}}= (p+1)\partial_{[\mu_1} B_{\mu_2\mu_3 \cdots \mu_{p+1}]},
\end{equation}
and the square brackets indicate complete anti-symmetrisation. 

The EHS is obtained by working in the gauge 
\begin{equation}
\label{ge}
B_{0\mu_1\cdots \mu_{p-1}}=0.
\end{equation}
The resulting constraints, analogous to the Gauss law, is 
\begin{equation}
\label{eomp}
\partial^i  H_{0  i \mu_1 \cdots \mu_{p-1}}=0.
\end{equation}
The different superselection sectors, for the electric centre choice  are therefore specified by the value for the normal component of the electric field,
$H_{0 n\mu_1\cdots \mu_{p}}\equiv H_{ 0 i \mu_1 \cdots \mu_{p}}n^i $.

The resulting classical term is then determined by the two-point function of the normal  component. For a planar boundary, which we continue to denote as 
$z=0$ we get
\begin{equation}
\begin{aligned}
\label{pf2pt}
\langle H_{0 z i_1 i_2 \cdots i_{p-1}} (\mathbf{k}_\parallel) H_{0 z j_1 j_2 \cdots j_{p-1}} (\mathbf{k}'_\parallel) \rangle \sim \delta^{(d-1)}({\mathbf{k}_\parallel+\mathbf{k}'_\parallel} )&\,
k_\parallel^2 \pqty{ \delta^T_{i_1[j_1} \delta^T_{i_2 j_2} \cdots \delta^T_{i_{p-1} j_{p-1}]}  }  \\
&\times \int \frac{\md k_z}{\sqrt{k_z^2+k_\parallel^2}} e^{\mi k_z \Delta}.
\end{aligned}
\end{equation}
Here 
\begin{equation}
\label{defdt}
\delta^T_{ij}\equiv \delta_{ij}-{\pqty{k_{\parallel }}_i \pqty{k_{\parallel }}_j \over k_\parallel^2}
\end{equation}
is the delta function transverse to the spatial momentum along the boundary,   ${\mathbf k_\parallel}$, and the square brackets in the product of the delta functions indicates complete anti-symmetrisation with respect to the second indices $j_1, j_2, \cdots j_{p-1}$.

We see that the two-point function, in general, involves the Laplacian for a $(p-1)$-form living on the boundary which is $(d-1)$-dimensional. 
The contribution is once again dominated by short distance modes and logarithmically dependent on the cut-off $\Delta$. 
As a result, we see that the contribution of the classical piece drops out of the mutual information or relative entropy.\footnote{As in the gauge field case, the logarithmic dependence on $\Delta$ is true for modes with $k_\parallel \ll {\Delta^{-1}}$. However, using as argument analogous to eq.(\ref{addtwo})-eq.(\ref{intdd}), it also follows the contribution of modes with $k_\parallel \sim {\Delta^{-1}}$ to the classical piece in the mutual information and relative entropy  drops out.}

Let us also note that when we work with a region which is not the half space ($z<0$ above) and the boundary  is no longer planar the logarithmic divergence will be cut off by the size of the region of interest. Also, we need to be 
more careful about zero modes in general.

\section{Entanglement and Dualities}\label{sec-duality}
In this section, we study some aspects of the duality which relates a theory with a $p$-form gauge potential   in $d+1$ dimensions to its dual which has a $(d-p-1)$-form potential.
This is a generalisation of electric-magnetic duality in $3+1$ dimension for a gauge field, and  is also  closely related to the Kramers-Wannier duality on the lattice. 
Our interest will be to study how the different choices for the centre of the algebra of observables transform under this duality and the accompanying changes in the entanglement entropy and related quantities. In particular, we will consider  two choices of centres for the $p$-form theory, called the electric and magnetic centres and study how they map under duality. We will find that the magnetic centre choice  for a region $R$ in the $p$-form case  maps to an algebra of observables in the dual theory which is closely related but not identical  to the electric centre of the dual $(d-p-1)$-form theory in a suitable region ${\tilde R}$ of the dual lattice. 
In the continuum limit, these differences become unimportant for ultraviolet insensitive quantities, like the relative entropy and mutual information, as was mentioned above, see also \cite{Casini:2013rba}, and the two dual theories therefore agree.

We will work on a  spatial lattice in the Hamiltonian formulation \cite{Kogut:1974ag}. This is related  to the choice of gauge eq.(\ref{ge}) in the continuum. Physical states satisfy the analogue of the Gauss law constraints eq.(\ref{eomp}) on the lattice. 
We start with considering a $U(1)$ gauge field in various dimensions and then extend the discussion to general $p$-forms.

We should note here that the idea of carefully dualising algebras under Kramers-Wannier duality has appeared before in the literature \cite{Casini:2014aia,Radicevic:2016tlt,Radicevic:2018okd,Campiglia:2018see}, but we apply it for a somewhat different purpose.
Specifically, our purpose is to understand what calculation in the dual theory corresponds to the EHS calculation in the original theory.
In particular, our work sheds some light on the calculation of \cite{Donnelly:2016mlc}, in which the authors calculated using the replica trick the difference between EHS calculations in the two dual theories. The ``entanglement anomaly'' of \cite{Donnelly:2016mlc} is roughly
\begin{equation}
  \Delta S_{\mathrm{DMW}} = S_{\text{electric centre}} - S_{\text{dual electric centre}} = S_{\text{electric centre}} - S_{\text{magnetic centre}},
  \label{eqn:dmw-interpretation}
\end{equation}
i.e., the difference between the electric centre and magnetic centre calculations on the same side of the duality.
For a more precise statement, see below.

\subsection{Gauge Field in $2+1$ Dimensions}\label{subsec:2+1}
\begin{figure}
\centering
\includegraphics[scale=0.65]{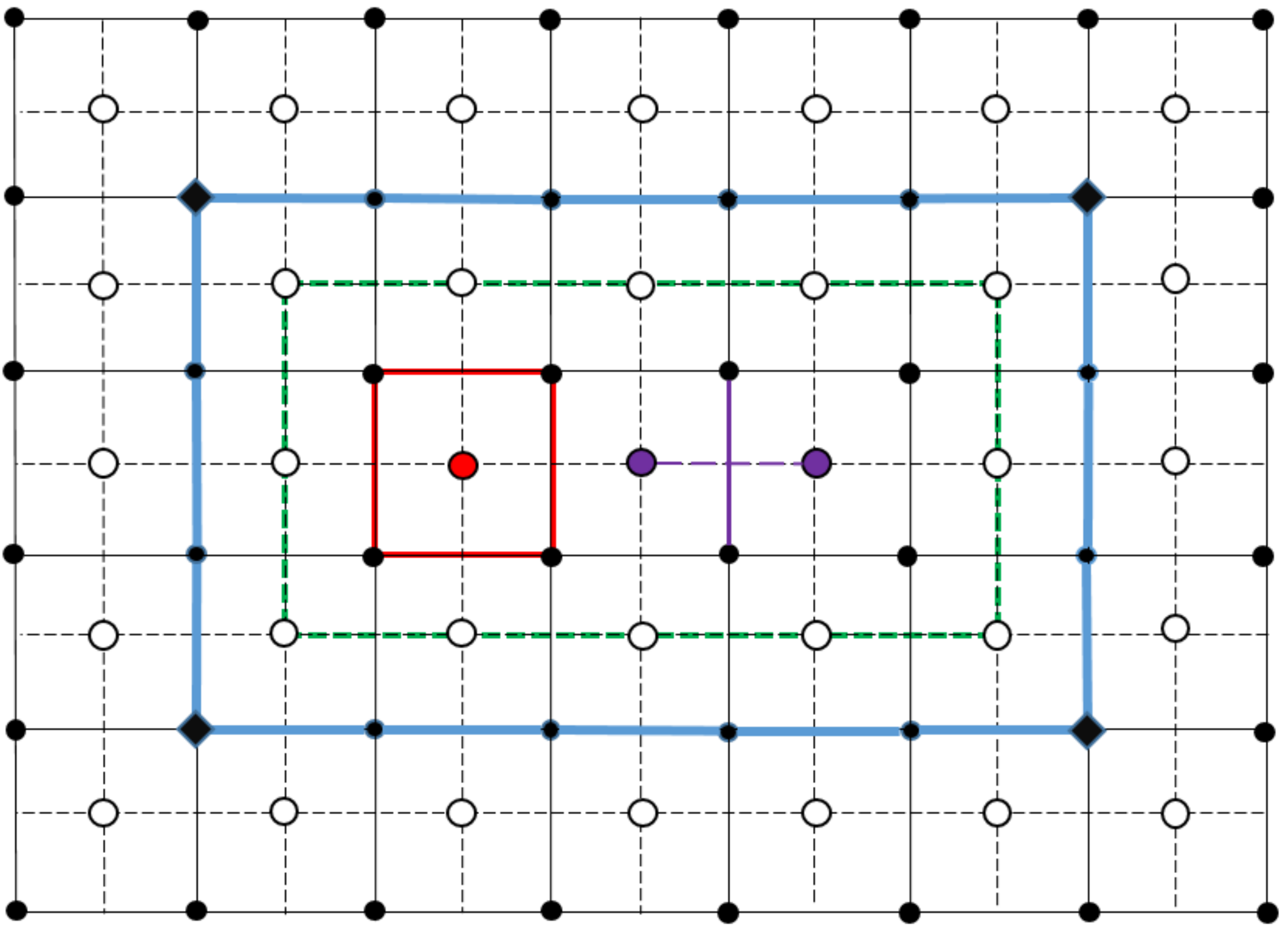}
\caption{The two-dimensional lattice for the  gauge theory (shown with solid lines) and the dual lattice for the scalar theory (shown with broken lines). Under duality, the electric operator along the solid violet link is mapped to the difference of $\phi$'s at the two violet vertices and the  Wilson loop along the  boundary of the  plaquette  shown in solid red is mapped to $\pi$ at the centre of the plaquette shown in red. The blue rectangle encloses the region $R$, with the four atypical boundary points at the edges shown as diamonds. The green rectangle encloses the dual region $\tilde R$.}
\label{fig:2d-lattice}  
\end{figure}

For simplicity, we start with the case of the $U(1)$ gauge theory in $2+1$ dimensions. We take the spatial lattice to be square. The degrees of freedom are link variables $\theta_l$ valued in $U(1)$. The gauge invariant observables are electric operators,
$E_l$,  which are conjugate to the link  variables  and Wilson loop  operators, which are defined for a   plaquette and given by  $\exp(\mi \sum _l \theta_l)$, where the sum is over oriented links along the boundary of the plaquette.
We will be  interested in the entanglement of a rectangular region $R$ shown in Fig \ref{fig:2d-lattice}. 
The electric centre  arises when we keep all gauge invariant operators in $R$ in the algebra and corresponds to specifying the normal component of the electric field for each boundary point. For the four  atypical  boundary points where more than one boundary edge meet, the sum of the electric fields along the two boundary edges meeting at the point   lie in the centre. By Gauss' law, this is equivalent to specifying the centre element  by the value of the sum of the electric fields along the outward directed links emanating from the boundary point.  
The magnetic centre is obtained  after one  deletes all the electric link operators for the boundary links from the algebra. It is easy to easy that the resulting algebra now has only one operator in the resulting magnetic centre which corresponds to the Wilson loop
along the boundary of the region.

 The dual lattice  is obtained by replacing each link of the original lattice with a dual link, in the standard fashion as shown in Fig \ref{fig:2d-lattice}. This exchanges points of the original lattice  with an elementary plaquette in the dual which surrounds the point and vice-versa. The dual theory lives on the dual lattice and  is that of a free scalar.
 The gauge invariant operators discussed above map as follows.
 The Wilson loop along an elementary plaquette surrounding the point $i$ in the dual is 
 \begin{equation}
 \label{wl}
 W=\pi(i),
 \end{equation}
 where $\pi(i)$ is the operator which is the conjugate momentum for the  scalar operator located at vertex $i$, $\phi(i)$. 
 The electric field along link $l$ is 
  \begin{equation}
 \label{elep}
 E_l\leftrightarrow \Delta \phi =\phi(i+1) -\phi(i),
 \end{equation}
 where $\phi(i)$ denotes the value of the scalar at site $i$ and $\phi(i+1)$ at the site $i+1$ which is displaced by one lattice unit orthogonal to the direction of the link $l$. See Fig.\ref{fig:2d-lattice}.
 
 The algebra of observables for the region of interest in the scalar case includes all operators  $\phi(i), \pi(i)$ with $i \in {\tilde R}$ where ${\tilde R}$ is the dual region shown in Fig \ref{fig:2d-lattice}. It is easy to see that this  maps to the magnetic  centre for the gauge theory with one small difference. The operator $\sum_i \phi(i)$, with the sum being over all points in ${\tilde R}$, is  absent in the gauge theory case. Its absence gives rise to a non-trivial centre which, as per the discussion above for the magnetic centre, is given by the Wilson loop which measures the total  magnetic flux threading $R$. This operator is dual to $\sum_{i\in {\tilde R}} \pi(i)$. Once the operator $\sum_{i\in {\tilde R}}\phi(i)$ is also present in the algebra, the centre becomes trivial. 

 Despite this difference, the entanglement of ground state is the same in the two algebras, as long as the theory is on a finite lattice.
 The reason is that the scalar theory has a global shift symmetry
 \begin{equation}
   \phi(i) \to \phi(i) + \alpha,
   \label{eqn:global-shift-symm}
 \end{equation}
 generated by the total conjugate momentum
 \begin{equation}
   \Pi^{T} = \sum_{i \in L} \pi(i).
   \label{eqn:total-pi}
 \end{equation}
 On a finite lattice, this symmetry is not spontaneously broken, and  the ground state is an eigenstate of  $\Pi^{T}$  with zero total momentum. 

 Since we can write $\Pi^{T}$ as a sum of total inside and outside conjugate momenta,
 \begin{equation}
   \Pi^{T} = \Pi_{\mathrm{in}}^{T} + \Pi_{\mathrm{out}}^{T},
   \label{eqn:total-pi-dec}
 \end{equation}
 the reduced density matrix must commute with $\Pi_{\mathrm{in}}^{T}$ and therefore must be block-diagonal in any eigenbasis of $\Pi_{\mathrm{in}}^{T}$.
 Because of this, the reduced density matrix does not change when the operator $\sum_{i \in \tilde{R}} \phi(i)$, which is conjugate to $\Pi_{\mathrm{in}}^{T}$, is deleted from the algebra.
 This means that the reduced density matrix, and therefore the entanglement spectrum, is the same as that of the magnetic centre algebra which is the algebra of the region without the operator $\sum_{i\in \tilde{R}} \phi(i)$ whose removal makes no difference.

  Let us end with a few comments. First, the argument above showing that the density matrices of the two algebras which differ by the inclusion of the operator 
  $\sum_{i\in {\tilde R} }\phi(i)$ are the same will be true in any eigenstate of the total momentum, $\Pi^{T}$.
  In contrast, the argument will  not apply for an infinite lattice since  the global shift symmetry is spontaneously broken  in this case and the ground state is no longer an eigenstate of $\Pi^T$.

  Second, it was found in  \cite{Donnelly:2016mlc} that, for theories on compact manifolds, the EHS entanglement for the gauge theory agrees with the scalar entanglement
  (with a suitable identification of the UV regulator). 
  The EHS entanglement includes a non-extractable piece which scales like  the boundary  area and  diverges in the continuum limit. In contrast, for the scalar, the full result for the entanglement is, of course, extractable. As a result,  the extractable entanglement in the two cases will not agree.
  Let us emphasise that this is not a contradiction since we are comparing two different algebras and the operations which can be carried out during quantum distillation or dilution depend on the operators at one's disposal. It is nonetheless interesting that the full entanglement does agree in the two cases. We also note that the scalar case
  and the magnetic centre only differ by  one operator, as noted above, and this difference is not expected to give rise to a differing  area law contribution for the magnetic centre and scalar  cases.

  Also, we have not been careful about the distinction between the compact and non-compact $U(1)$ theories above. It is well known that the compact $U(1)$ theory  maps to a compact boson (see, for example, \cite{Agon:2013iva} for a careful discussion). The momenta $\pi(i)$ are well defined operators in the compact case, but instead of the field $\phi(i)$ we can work with the phase $e^{\mi \phi}$ which is well defined (for $\phi$ having periodicity $2\pi$). The duality map can then be easily worked out and the discussion above can be extended in a precise manner  for the compact case as well.

 \subsection{Gauge Field in $3+1$ Dimensions}\label{subsec:3+1}

 Let us next consider a $U(1)$ gauge theory in $3+1$ dimensions. 
 Its dual is also   a $U(1)$ gauge theory. Under duality,  the electric and magnetic fields are exchanged with each other,
 \begin{equation} 
 \begin{aligned}
 \label{dualem}
 \mathbf{E} &\to   \mathbf{B},\\
 \mathbf{B} &\to  -\mathbf{E}.
 \end{aligned}
 \end{equation}
  
  The Gauss law and the Bianchi identity are exchanged  under this map. 
  On the lattice, the magnetic field    actually refers to  the value of the Wilson loop around an elementary plaquette,
 and the electric field to the corresponding electric operator on a link.    Under duality, links are exchanged with plaquettes on the dual lattice and this exchanges the electric link and Wilson loop operators, in a lattice version of eq.(\ref{dualem}). The electric field operators and the Wilson loop operators constitute a complete set of gauge invariant observables. In our subsequent discussion, we will sometimes be a little loose and refer to the electric link operators and Wilson loop operators on the lattice as electric and magnetic fields. 

  \begin{figure}
\centering
\includegraphics[scale=0.6]{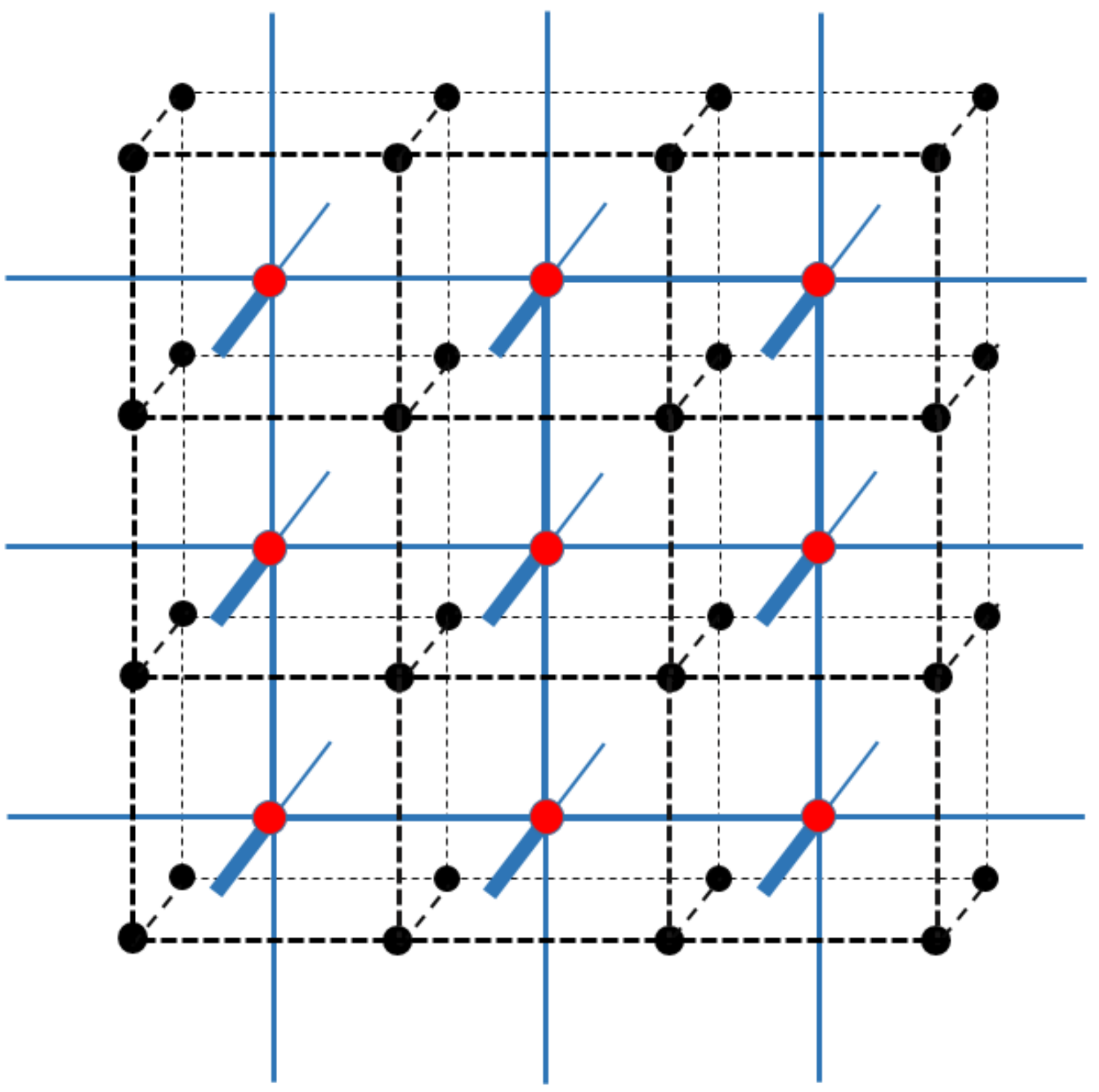}
\caption{A boundary face in the three dimensional case.  Wilson loops  that lie in the magnetic centre are shown as dashed lines on the boundary face. Their duals are electric link operators along the outward directed normals, shown as blue links emanating from the vertices of the dual lattice denoted in red.}
\label{fig:3d-lattice}  
\end{figure}
 The electric centre is obtained when we keep  all the gauge invariant operators in the  region of interest in the algebra. Due to the Gauss law constraint, the centre elements can be thought of as  corresponding to the normal component of the electric field on the boundary.

 The magnetic centre is obtained,  as in the $2+1$-dimensional case, by removing all electric link operators on the boundary from the algebra. The resulting algebra then has as its centre the Wilson loop  operators along all the plaquettes that lie on  the boundary  faces. Under duality, these are exchanged with electric link operators. For the region $R$ of interest in the original theory, we can define a dual core region ${\tilde R}$ obtained by displacing all boundary lattice points half a lattice unit inwards along the normal. The electric link operators dual to the Wilson loops in  the magnetic centre of the original theory lie along links  which go  from the boundary points of ${\tilde R}$ outwards along the normal direction. In Fig \ref{fig:3d-lattice}, we show one boundary face and the plaquette operators in the original lattice which are  in the magnetic centre and lie on this face along with their dual electric link operators. 
 
 The dual description of the magnetic centre case can therefore be thought of as follows. 
 We augment the set of all gauge invariant operators in the region ${\tilde R}$ with the additional electric link operators along links which extend from boundary points outward along the normal. This extended algebra then has the additional electric operators as its centre. For a typical boundary point of ${\tilde R}$, where no boundary faces meet, this additional electric operator is related to the other  electric link operators emanating from the boundary point, which are already present in the algebra, by the Gauss law constraint. Therefore, the addition of this operator to the algebra actually does not change it.  
 However, for boundary points where more than one face meet  the additional electric link operators added in this way do enlarge the algebra, since  Gauss' law only relates the sum of these outgoing electric link operators  to the electric link operators already present in ${\tilde R}$.  
 
 In this way, we see that the magnetic centre  case in the original theory is related to the    electric centre case in the dual theory by the addition of some additional electric link operators, which in the continuum limit we expect to be the same as the electric centre case.  Conversely, the electric centre  case of the original theory, with the addition of some extra  electric link operators at boundary points where more than one face meet, maps to the magnetic centre case of the dual theory.

 We end with the following comment. In the continuum limit, an alternative is to  just choose a different latticisation where the boundary does not have `corners;' for example, if the boundary is a sphere, we can choose the links in the lattice to consist of radial lines and, at a particular radius, latitudes and longitudes of the sphere.
 While this lattice is extremely odd at the origin, the main point is that in this lattice the electric centre exactly dualises to the magnetic centre, thereby tying in with the expectation that the additional operators discussed above will not make a significant difference in the continuum limit.

  \subsection{$p$-Form Theory in $d+1$ Dimensions}\label{subsec:p}
   It is now easy to generalise the discussion for the general case of a $p$-form gauge potential in $d+1$ dimensions. This is dual to a $(d-p-1)$-form potential theory. 
   We denote  the components of the potential by $B_{\mu_1 \mu_2 \cdots \mu_p}$ and the field strength by $H_{\mu_1 \mu_2 \cdots \mu_{p+1}}$ and those of the dual potential and dual field strength by 
   ${\tilde B}_{\mu_1 \mu_2 \cdots \mu_{d-p-1}}$ and  ${\tilde H}_{\mu_1 \mu_2 \cdots \mu_{d-p}}$ respectively. On the spatial  lattice, the gauge invariant observables are the electric fields $H_{0 i_1i_2 \cdots i_p}$ which are  defined on a $p$-dimensional hypercube, and the magnetic fields are given by the oriented sum of $H_{i_1 i_2  \cdots i_{p+1}}$ along all the faces of a $(p+1)$-dimensional hypercube. 
   The Gauss law constraint in the lattice takes the following form: Consider  all $p$-dimensional hypercubes which share a vertex. Then the 
   sum of all electric fields which share $p-1$ directions in common on these hypercubes   must vanish.   The Bianchi identity is given as follows. Consider a $(p+2)$-dimensional hypercube which has $(p+1)$-dimensional hypercubes as its faces. Then the oriented sum of the magnetic fields along all these faces vanishes. The Gauss law constraint and the Bianchi identity are exchanged under duality.

  For a region $R$ which is now a  $d$-dimensional hypercube, the boundary faces are $(d-1)$-dimensional. The outward normal component of the electric field  at a boundary point corresponds to hypercubes which have one link extending from the boundary point along the outward normal and the remaining $p-1$ links  along the $d-1$ directions tangent to the boundary. The electric centre case arises when all gauge invariant operators present in $R$ are included in the algebra. Using Gauss' law, the centre can  then be thought of as the outward normal component of the electric field at all boundary points.     The magnetic centre is obtained by removing all electric operators which take values on hypercubes that lie along the boundary. It consists of all the magnetic field operators which take values on $(p+1)$-dimensional hypercubes lying on the boundary. 
  
  Consider now the magnetic centre in the dual theory.  The dual core region ${\tilde R}$ is as above given by the region in the dual lattice obtained by going half a lattice unit inwards from the boundary of $R$ along the normal. The magnetic centre operators of the $p$-form theory map to  outward normal components of the electric field at the boundary points of ${\tilde R}$ in the dual theory. Under the duality map, these operators are added to all the gauge invariant operators in ${\tilde R}$ to give the full algebra. Using Gauss' law, as in the case of the $U(1)$ theory in $3+1$ dimensions above, the additional operators actually only arise from boundary points of ${\tilde R}$ where more than one faces meet. 
  
    The one exception which needs to be handled separately is the case of a $(d-1)$-form gauge potential in $d+1$ dimensions. This is  analogous to case of a  gauge theory in $2+1$ dimensions, discussed above. Here the dual theory is that of a  scalar, and the magnetic centre case of the original theory maps to the situation where all operators in the scalar theory lying in the dual core region are included in the algebra, with the exception of $\sum_{i\in {\tilde R}} \phi(i)$. The centre has one element which in the total magnetic flux in $R$ or in dual description the total momentum in ${\tilde R}$, $\sum_{i\in {\tilde R}} \pi(i)$. Including the operator $\sum_{i\in {\tilde R}} \phi(i)$ gives rise to a trivial centre.  Using an argument  analogous to that  in \S\ref{subsec:2+1}, it follows that the addition of this operator does not change the density matrix and the entanglement entropy for a state which has a fixed value of the total momentum,
    $\sum_{i \in {\tilde L}}\pi(i)$.
    
\section{Conclusions}\label{sec:conclu}
In this paper, we have analysed some  aspects pertaining to the entanglement entropy for gauge theories. 
We argued that for the extended Hilbert space definition, or equivalently the electric centre case, with entanglement entropy given in eq.(\ref{resee}), the probability to 
be in the different superselection sectors, $p_i$,  is determined by the two-point function of the normal component of the  electric field along the boundary -- more specifically, eq.(\ref{twopac}) in the Abelian and eq.(\ref{tptym}) in the non-Abelian cases. We find that the two-point function is   dominated by modes which carry high momenta in the direction normal to the boundary. These modes give rise to a UV divergent contribution to the two-point function.  As a result, in the continuum limit,  the contribution from the first two terms  in eq.(\ref{resee}) will drop out  from the mutual information of two disjoint regions and  the relative entropy between two states which have only finite energy excitations about the vacuum.

 It is worth noting that our conclusions  also apply to theories with matter. 
Strictly speaking,  we assume that the  theories are weakly coupled at short distances. In the non-Abelian case this would be true for asymptotically free theories. Our  conclusions could  be altered if the UV behaviour is different, from example  if the behaviour is  governed by a fixed point which changes the anomalous dimensions of the electric field operators sufficiently from the free field case. 
We also note   that in the Abelian case, the high momentum modes give rise to a logarithmic divergence in the two-point function of the normal component of the electric field, this divergence is universally present irrespective of dimension.\footnote{Strictly speaking, this is true for modes with $k_\parallel \ll {\Delta^{-1}}$, eq.(\ref{twopaba}).}

Let us also mention that the  behaviour of $2+1$-dimensional Maxwell theory has been studied numerically, \cite{Casini:2014aia}. It was found that the the dependence on the centre does indeed drop out in the continuum limit. It was also argued quite generally   using the expectation that the relative entropy and mutual information for ``thin regions'' should vanish in the continuum limit   that the contribution of the classical term  to the relative entropy or the mutual information would  drop out in this limit \cite{Casini:2013rba}.
 Some evidence for this was  also obtained numerically, \cite{Casini:2014aia}. Our analysis confirms these expectations.

In the more general case of  an Abelian  $p$-form gauge potential theory in $d+1$ spacetime dimensions, we find that the corresponding probability distribution is determined by the Laplacian for a $(p-1)$-form living on the $(d-1)$-dimensional spatial  boundary of the region of interest. The high momentum modes dominate the behaviour of the relevant  two-point function  in this more general case as well and give rise to a  similar logarithmic divergence in the   two-point function.
As a result, the classical term in the entanglement, $-\sum_i p_i \log p_i$, drops out in the continuum limit from    the mutual information and  the relative entropy  which only receive a contribution from the quantum piece in the entanglement entropy. We have not been  very careful about studying  the zero modes of the $(p-1)$-form theory.  See \cite{Donnelly:2016mlc} for an important discussion in this regard; 
see also the  comments at the end of \S\ref{subsec:abel}, where we discuss how in Maxwell theory the zero modes can sometimes lead to  contributions in the mutual information or relative entropy arising from the classical term.

Finally, we have also briefly studied some aspects of electric-magnetic duality on the lattice. We find that the magnetic centre for a $p$-form in $d+1$ dimensions maps to the electric centre case for the dual $(d-p-1)$-form, with the addition of a few more operators to the algebra. In the continuum limit, these extra operators will not make a difference 
while computing the mutual information or the relative entropy and the electric and magnetic theories will therefore agree in the results  we obtain for these quantities.

It will be worth extending our discussion for gravity in the linearised limit. For some existing  discussion, see \cite{Donnelly:2016auv,Speranza:2017gxd,Geiller:2017whh,Camps:2018wjf}. 
The classical term can be thought  of as arising due to gauge transformations with support on the boundary of the region of interest. 
Elucidating the nature of these boundary degrees of freedom in the gravity case in more detail, and the resulting classical term, both for a planar boundary with a Rindler horizon, and at asymptotic null infinity would be quite interesting. It would also be interesting to explore how these considerations might be related to proposals for soft hair on a black hole horizon, \cite{Hawking:2016msc,Hawking:2016sgy,Haco:2018ske}. 

\acknowledgments

We thank Horacio Casini, William Donnelly, Shamit Kachru, Suvrat Raju, V Vishal and Aron Wall for insightful discussions.  SPT thanks the organizers of the “It From Qubit” workshop (January 4-6, 2018) in Bariloche, Argentina for support, where some of the results were presented.  This research was supported in part by the International Centre for Theoretical Sciences (ICTS) during a visit (by UM and RMS) for participating in the programme Kavli Asian Winter School (KAWS) on Strings, Particles and Cosmology 2018 (Code: ICTS/Prog-KAWS2018/01). UM thanks the “Infosys Foundation International Exchange Program of ICTS” for support in participating in Strings 2018 held in Okinawa, Japan. UM gratefully acknowledges support from the Simons Center for Geometry and Physics, Stony Brook University for participating in the 2018 Simons Summer Workshop while this work was in progress. UM and SPT acknowledge support from ICTS during a visit for participating in the programme “AdS/CFT at 20 and Beyond”. We thank the DAE, Government of India, for support. We acknowledge support from the Infosys Endowment for Research on the Quantum Structure of Spacetime. SPT also acknowledges support from the J. C. Bose fellowship of the DST, Government of India. Most of all, we are grateful to the people of India for generously supporting research in String Theory. 

\bibliographystyle{JHEP}
\bibliography{ref}
\end{document}